\documentstyle[12pt,epsf]{article}
\begin{document}

\begin{center}
{\bf Proton and neutron charge form factors in soliton model\\[0pt]
with dilaton-quarkonium field. }\\[10mm]
{\sc Lassila K.E.$^{1}$}	\\
{\it International Institute of Theoretical and Applied Physics,	
Ames, USA}	\\[5mm]
{\sc Magar E.N.$^{2}$}	\\
{\it Laboratory of Computing Techniques and Automation,	\\
Joint Institute for Nuclear Research, Dubna, Russia}	\\[5mm]
{\sc Nikolaev V.A.$^{3}$}	\\
{\it Science Applications and Human Resources Agency,	\\
London, UK}	\\[5mm]
{\sc Novozhilov V.Yu.$^{4}$}	\\
{\it S-t Petersburg State University, 	
St.Petersburg, Russia}	\\[5mm]
{\sc Tkachev O.G.$^{5}$}	\\
{\it Institute of Physics and Information Technologies,	\\
Far East State University, Vladivostok, Russia}	\\[10mm]
{\it E-mail: 
	$^{1}$klassila@iastate.edu 
	$^{2}$magar@hepalpha1.physics.smu.edu 
	$^{3}$chiralnet@myfreeoffice.com 
	$^{4}$novozhil@heps.phys.spbu.ru 
	$^{5}$tkachev@ifit.phys.dvgu.ru}
\end{center}
\setcounter{footnote}{5}
\footnotetext{This work has been partially supported by grant "Universities of Russia"
N 02.01.22}
\setcounter{footnote}{1}

\begin{abstract}
Nucleon electromagnetic form factors are considered in the framework  of the
generalized Skyrme model with dilaton-quarkonium field. In our recent
publication  we have got big discrepancies between calculated form factors
and dipole approximation formula. Here we have reasonably good accordance
between them in finite impulse region after vector meson dominance have been
taken into account. Omega and Rho -meson have been included into only hadron
structure of the photon.
\end{abstract}

\newpage

\section{Introduction}
\indent

The Lagrangian appropriate for a generalized Skyrme model in 
the leading classical field approximation
yields chiral soliton solutions.  This solitons are associated with
the nucleon and will be used for obtaining spatial structure information
on baryons.  Effectively, then, the baryon or nucleon form factors can
be extracted from the soliton model equations of motion.  Experimentally
measured form factors can be compared with these soliton predicted 
spatial densities to test the model.

The present paper includes discussion of our recent work, extensions,
and calculations of the nucleon electromagnetic form factors in the
generalized Skyrme model.  This model involves the theoretical
description of the dilaton-quarkonium scalar field and shows its
importance in the description of soliton dynamics.  We use 
"dilaton-quarkonium" scalar field to indicate the way we 
subdivide the gluon condensate in the calculation.  This generalized
model reproduces the experimental value of the nucleon mass, the input 
being the experimental value of the pion decay constant and the 
theoretically derived value of the Skyrme constant, 
$e = 2\pi$.\cite{NuovoCimento}.  The naive, straightforward calculation 
of the electromagnetic form factors has shortcomings: the  
values of $F_{\pi}$ and $e$ give too small a nucleon size and 
the calculated curves do not give the approximate dipole form factor
values. 

The generalized Skyrme model under consideration follows the formulation
of Andrianov {\it et al.} \cite{Andrianov2}  This approach uses the
framework of the joint chiral and conformal bosonization of the QCD
Lagrangian, including chiral and scalar dilaton-quarkonium fields.  
In such a model the properties of the topological solitons are 
dramatically changed in numerical value from those in the original 
Skyrme model.  Several authors 
have introduced an additional scalar field to the Skyrme model for 
different motivational reasons.  For example, Riska and 
Schwesinger\cite{Riska88} appear to be the first to 
investigate the isospin independent part of the nucleon-nucleon
spin-orbit interaction when a scalar field is added.  A number of papers
studied the effects a scalar $\sigma$ meson would have by introducing 
it as a gluon condensate.\cite{Gomm1},\cite{Gomm2},\cite{Jain}, and also 
related, \cite{Andrianov2} and \cite{Andrianov3}.  The purely
theoretical and convincing reason is that with the introduction of a
scalar field, the conformal anomaly, one of the distinctive features of
the QCD Lagrangian, is reproduced.  In the SU(2) sector of this
Lagrangian, one can construct an effective theory which reproduces the
conformal anomaly in the framework of the effective Lagrangian method,
introducing a field corresponding to scale invariance.  As shown 
in \cite{Lacombe}, \cite{Yabu},and \cite{Mashaal}, it leads to
the necessary strong attraction at intermediate internucleon distances.
In such
an approach the starting point is the fermion integral over quark
fields, in the low energy regime of QCD.  The integral is specified by
the finite mode regularization scheme with a cut-off that also plays the
role of a low energy boundary.  Performing the joint chiral and
conformal bosonization on this integral leads to an effective action for
chiral $U(x)$ and dilaton $\sigma(x)$ fields.  This Lagrangian favors
the linear sigma model in terms of the composite field
$U(x)exp(-\sigma(x))$.  The resulting effective
Lagrangian\cite{Andrianov2}, generalizing the original Skyrme Lagrangian
is 
        \begin{eqnarray}
        L_{eff}(U,\sigma) &=& {F_\pi ^2\over 4}exp(-2\sigma )Tr[\partial_\mu
        U\partial ^\mu U^+]+\nonumber\\ \cr
        &&{{N_fF^2_\pi }\over 4}(\partial_\mu\sigma )^2
        \exp(-2\sigma )+\nonumber\\ \cr
        &&+{1\over 128\pi^2}Tr[\partial_\mu U U^+,
        \partial_\nu UU^+]^2-\nonumber\\ \cr
        &&{{C_gN_f}\over 48}(e^{-4\sigma }-1+{4\over
        \varepsilon}(1-e^{-\varepsilon\sigma}))
        \label{c8}\end{eqnarray}
where the pion decay constant is taken as the experimental value, 
$F_\pi = 93 MeV$ and $N_f$ is the number of flavors.  The gluon
condensate, according to QCD sum rules, is 
$C_g=(300-400 MeV)^4$\cite{Novikov0}.  The first two terms are the
kinetic terms for the chiral and scalar fields and the third term, the
well-known Skyrme term.  The effective potential for the scalar field is
the result of an extrapolation\cite{Andrianov2} of the low energy potential
to high energies by use of a one-loop-approximation to the Gell-Mann Low
QCD $\beta$ - function.  The parameter $\varepsilon$ is determined by
the number of flavors $N_f$ as $\varepsilon={8N_f}/ (33-2N_f)$.

\section{The Nucleon}
\indent

In the baryon sector we choose the chiral field as the spherically
symmetric ansatz of Skyrme and Witten, {$U(\vec x)=exp[-i\vec \tau\vec 
n F(r)]$}, where ${\vec n}={\vec r} / |\vec r|$. It is convenient to
introduce a new field, $\rho (x)=exp(-\sigma (x))$.  Then, the mass
functional in dimensionless variables, $x=eF_\pi r$, has the form 
$M=M_2+M_4+V$, where 
        \begin{eqnarray}
        M_2 &=& 4\pi {F_\pi\over e}\int_{0}^{+\infty}dx[{N_f\over 4}x^2
        (\rho\prime)^2 +\rho ^2({x^2(F')^2\over 2}+sin^2F)]\ ,
        \label{c9}\\ \cr
        M_4 &=& 4\pi{F_\pi\over e}\int_{0}^{+\infty}dx({sin^2F\over {2x^2}} +
        (F\prime ) ^2)sin^2F\ ,
        \label{c10}\\ \cr
        V &=& 4\pi {F_\pi\over e}D_{eff}\int_{0}^{+\infty}dx x^2[\rho^4-1+
        {4\over\varepsilon}\cdot (1-\rho ^\varepsilon )]\ .
        \label{c11}\end{eqnarray} 

In the last equations , the same Skyrme parameter value, $e = 2\pi$, is 
used.  The contribution of the potential to the mass is determined by
the factor $D_{eff}=C_gN_f/48e^2F_\pi^4$.  The mass functional leads to
a system of equations for the profile functions $F(x)$ and $\rho (x)$, 
where a prime is used to denote the derivative with respect to $x$, 
        \begin{eqnarray}
        F''[\rho^2x^2+2sin^2F]+2F'x[x\rho\rho '+\rho ^2]+(F')^2\cdot
        sin(2F)-\nonumber\\ \cr
        -\rho ^2\cdot sin(2F)-sin(2F)\cdot sin^2F/x^2 =0\ ,
        \label{c12}\end{eqnarray}
        \begin{eqnarray}
        {N_f\over 2}x[x\rho ''+2\rho ']-2\rho [{x^2(F')^2\over 2}+sin^2F] -
        \nonumber\\ \cr
        - 4D_{eff} \cdot [\rho ^3-\rho^{\varepsilon -1}] x^2= 0\ ,
        \label{c13}\end{eqnarray}
At small distances, $F = \pi N-\alpha x$ and $\rho = \rho (0)+\beta
x^2$, with $\rho (0) \neq 0$.  For large $x$, these functions behave as 
$F(x)\sim a/x^2$, and $\rho(x)\sim 1-b/x^6 + \dots$.

According to the virial theorem,\cite{Nikolaev109} the contributions of
the individual terms of the mass functional to the energy of
the system must obey the condition, 
\begin{eqnarray} M_4 
- M_2 -3V = 0 \ , \label{c13a}\end{eqnarray} 
which can be used to control the accuracy of the numerical solution of
the system.  There are nontrivial equations between the numbers 
 $\alpha$ and $\beta$, $a$ and $b$, 
 \begin{eqnarray} b &=&
        {1\over 2} a^2/D_{eff} \ , \label{c13b}\\ \cr \beta &=&
        \left[\rho(0)\alpha^2+{4\over 3}\left(\rho^3(0)-1\right)
        D_{eff}\right]/N_f\ .
        \label{c13c}\end{eqnarray}
The choice of boundary conditions ensures a finiteness of the mass functional for a
given value of the topological charge $B = N$.
Performing canonical quantization of the rotational degrees of freedom 
with the collective variable method,\cite{Witten} one obtains for the
nucleon mass,
        \begin{eqnarray}
        M_B= M+S(S+1)/(2I)\ ,
        \label{c14}\end{eqnarray}
where the moment of inertia is 
        \begin{eqnarray}
        I={8\pi\over 3}(F_\pi e)^{-3}\int_{0}^{\infty}dx\ sin^2[\rho^2x^2 +
        (F')^2 x^2+sin^2F]\ .
        \label{c15}\end{eqnarray}
Some numerical results are presented in Table \ref{StatProp1},
\begin{table}[ht]
\vspace{0.5cm}
{ \hfill
\begin{tabular}{| l | l | l |}
\hline
               & Present work & Original Model\\
\hline
 $M$		&   839 MeV	& 1098 MeV	\\
 $<r^2>^{1/2}$	&   0,44 Fm	&  0,42 Fm	\\
 $M_B$		&   1026 MeV	& 1288 MeV 	\\
\hline
\end{tabular}    
\hfill    }
\caption{Static properties ($N_f = 2$) in the generalized
Skyrme model with $F_\pi =93 MeV,\ e=2\pi ,\ C_g=(300 MeV)^4$.  Also, 
the results following from the original Skyrme model are given for
comparison purposes.}
\label{StatProp1}
\end{table}
where the soliton mean square radius of the corresponding baryon 
$<r^2_{IS}>$ density distribution is given
 \begin{eqnarray}
        <r^2_B>^{1/2}={1\over F_\pi e}\biggl\{-{2\over\pi}
\int_{0}^{\infty}dx
        x^2F' sin^2F\biggr\}^{1/2}\ .
        \label{c16}\end{eqnarray}

A discussion of partial restoration of chiral symmetry in this model is
given in Ref.(\cite{NuovoCimento}. The restoration appears as a large
deviation of $\rho(0)$ from its asymptotic value of $\rho(0)=1$.  The
dependence of the mass spectra on the gluon condensate in the
generalized Skyrme model was also discussed in \cite{NuovoCimento}.
 
\section{Form factors of charge distributions}
\indent

The nucleon electric and magnetic form factors, $G_E(q^2) $and 
$G_M(q^2)$ can be calculated from the electromagnetic currents, in
the Breit frame where the photon does not transfer energy. 
\begin{eqnarray}
<N_f({\vec q\over 2})| J_0(0) |N_i(-{\vec q\over 2})> &=& G_E(\vec 
q^2) \xi_f^+ \xi_i\ ,\nonumber\\
<N_f({\vec q\over 2})| \vec J(0) |N_i(-{\vec q\over 2})> &=&
{G_M(\vec q^2)\over 2M_N} \xi_f^+ i\vec\sigma\otimes\vec q\xi_i\ .
\label{F2}
\end{eqnarray}
\noindent
Here, $|N(\vec p)>$ is the nucleon state with momentum $\vec p$, 
$\xi_i$, $\xi_f$ and two component Pauli spinors, and $\vec q \equiv $ 
momentum transfer.

The isoscalar (S) and isovector (V) nucleon form factors are related to
those for the proton and neutron by 
\begin{eqnarray}
G^{{\rm p,\ n}}_{E,M}=G^{S}_{E,M} \pm G^{V}_{E,M}
\label{SVfactors}
\end{eqnarray}
These form factors are normalized to the respective charge and magnetic
moments by

\begin{figure}[t,h]
\epsfysize=10cm
{\hfill	\epsfbox{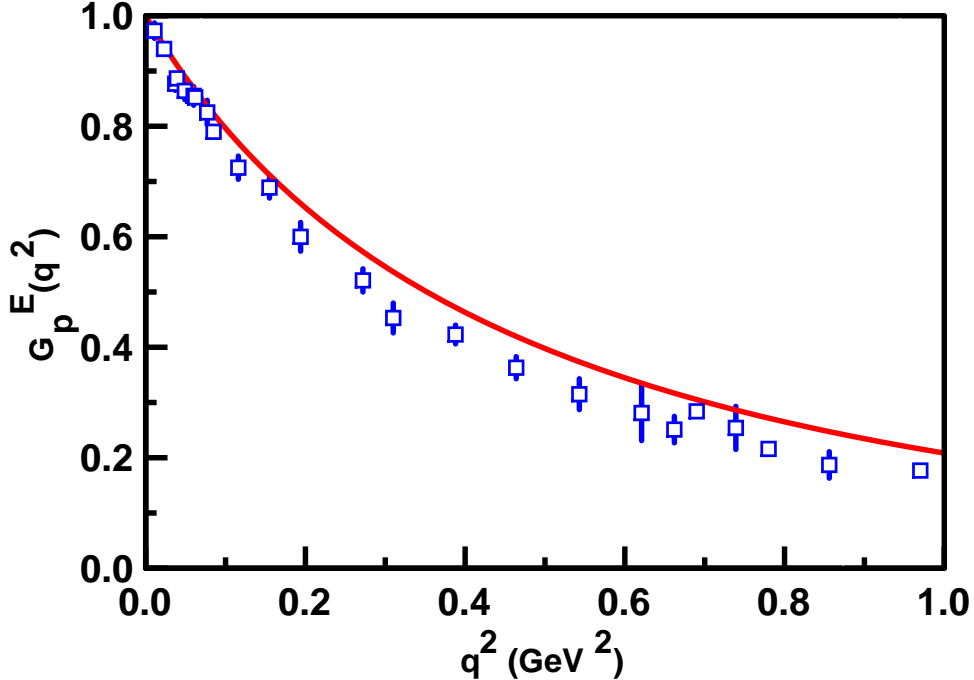}	\hfill}
\caption{Proton electric form factor as a function of $q^2$ in $GeV^2$
calculated for $F_\pi = 93$, $e=2\pi$, $N_f=2$, $C_g=(300 MeV)^4$, and 
$m_\pi=139$.  The experimental data shown come from} Ref.~\cite{Bartel}
\label{Proton}
\end{figure}

\begin{eqnarray} 
&&G^{\rm p}_{E}(0)=1 \qquad G^{\rm n}_{E}(0)=0	\cr
&&G^{\rm p}_{M}(0)\equiv \mu_p=2.79 \qquad G^{\rm n}_{M}= \mu_n=-1.91\ .
\label{Norm}
\end{eqnarray}
We remaked above on the smallness of the nucleon size as determined
by the baryon charge density distribution in the model with a
dilaton-quarkonium field.  

\begin{figure}[t,h]
\epsfysize=10cm
{\hfill	\epsfbox{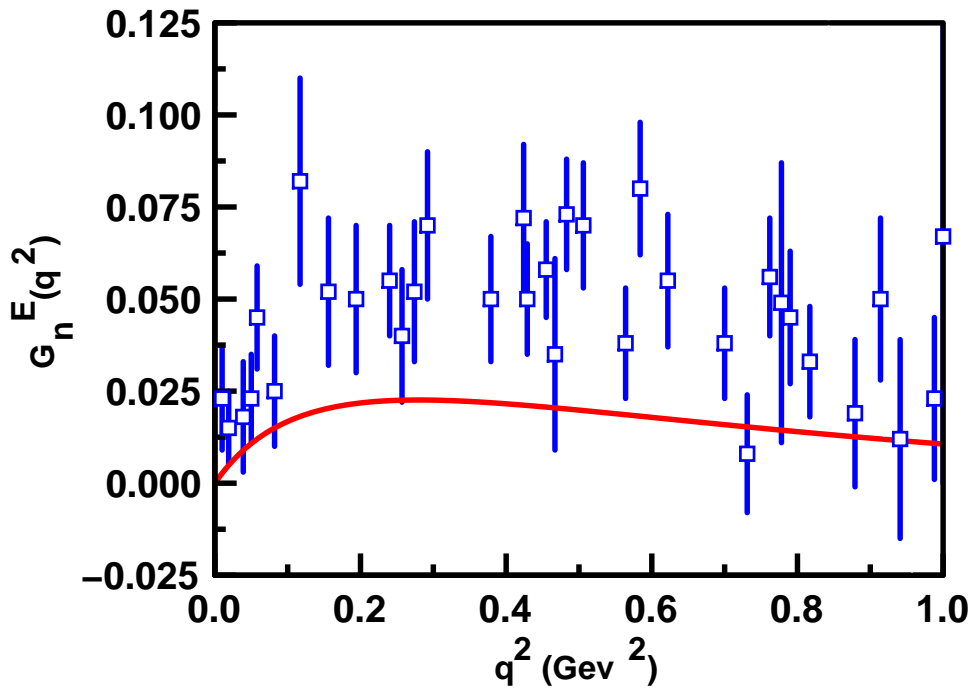}	\hfill}
\caption{Neutron electric form factor as a function of $q^2$ in $GeV^2$
calculated for $F_\pi = 93$, $e=2\pi$, $N_f=2$, $C_g=(300 MeV)^4$, and 
$m_\pi=139$.  The experimental data shown come from} Ref.~\cite{Bartel}.
\label{Neutron}
\end{figure}

Vector meson dominance means that the isoscalar photon sees 
$\omega$ meson ${\it structure}$, but not the isoscalar baryon density 
$B_0(r)$.

According to vector meson dominance, the isoscalar current is
proportional to the $\omega_{\mu}$-field, 
\begin{eqnarray}
J^{\mu}_{I=0}=-\frac{m^2_{\omega}}{3g}\omega_{\mu}(r)
\label{ICurrent}
\end{eqnarray}
and the corresponding charge form factor,
\begin{eqnarray}
G^{S}_{E}(\vec q^2)=-\frac{m^2_{\omega}}{3g}\int d^3r\exp{i\vec q\vec r}\omega(r)\ .
\label{IFactor}
\end{eqnarray}

The static $\omega (r)$ obeys the equation,
\begin{eqnarray}
(\nabla^2-m^2_{\omega})\omega (r)=\frac{3g}{2}B(r)=-\frac{3gF'(r)}{4\pi r^2}sin^2F(r)\ .
\label{EqForOmega}
\end{eqnarray}
From this equation, we obtain,
\begin{eqnarray}
G^{S}_{E}(\vec q^2)=-\frac{1}{2}\frac{m^2_{\omega}}{m^2_{\omega}+
\vec q^2}4\pi\int drr^2B_0(r)j_0(qr)\ .
\label{SFormFact}
\end{eqnarray}
Therefore, the effecive isoscalar nucleon density is equal baryon charge
density $B_0(r)$ times the $\omega$-meson propagator.  

\begin{figure}[t,h]
\epsfysize=10cm
{\hfill	\epsfbox{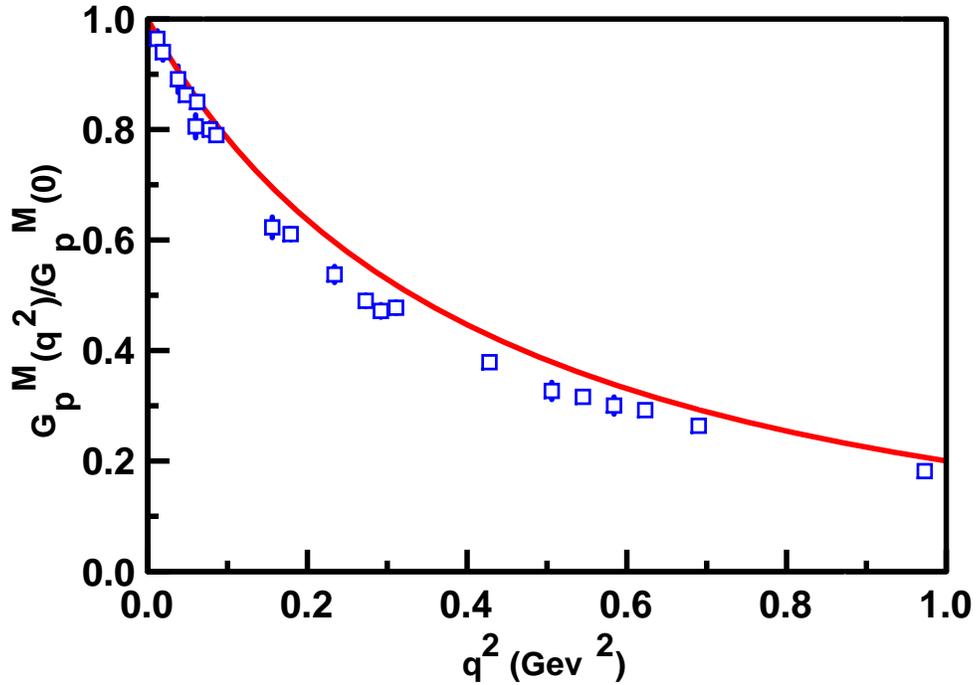}	\hfill}
\caption{Proton magnetic form factor as a function of $q^2$ in $GeV^2$
calculated for $F_\pi = 93$, $e=2\pi$, $N_f=2$, $C_g=(300 MeV)^4$, and 
$m_\pi=139$.  The experimental data shown come from} Ref.~\cite{Bartel}.
\label{ProtonM}
\end{figure}

The isovector electromagnetic formfactor has analogous structure,
\begin{eqnarray}
G^{V}_{E}(\vec q^2)=-\frac{1}{2}\frac{m^2_{\rho}}{m^2_{\rho}+\vec q^2}F^V_E(q^2)\ ,
\label{VFormFact}
\end{eqnarray}

In writing the propagators separately, as a factor 
multiplied into  $F^V_E$, $\omega$ and $\rho$, themselves have no
substructure or internal dynamics; the corresponding Skyrmion densities 
are considered as the sources of these $\omega$ and $\rho$ fields.  
Explicit considerations of the role of vector mesons in the
electromagnetic form factors in the $\sigma$ model has been given by 
Holzwarth\cite{HOLZ} and quantum corrections to the relevant baryon
properties in the chiral soliton models has been calculated.\cite{MEIER} 

The results of the present calculations are given in Figures 1 to 4.
\begin{figure}[t,h]
\epsfysize=10cm
{\hfill	\epsfbox{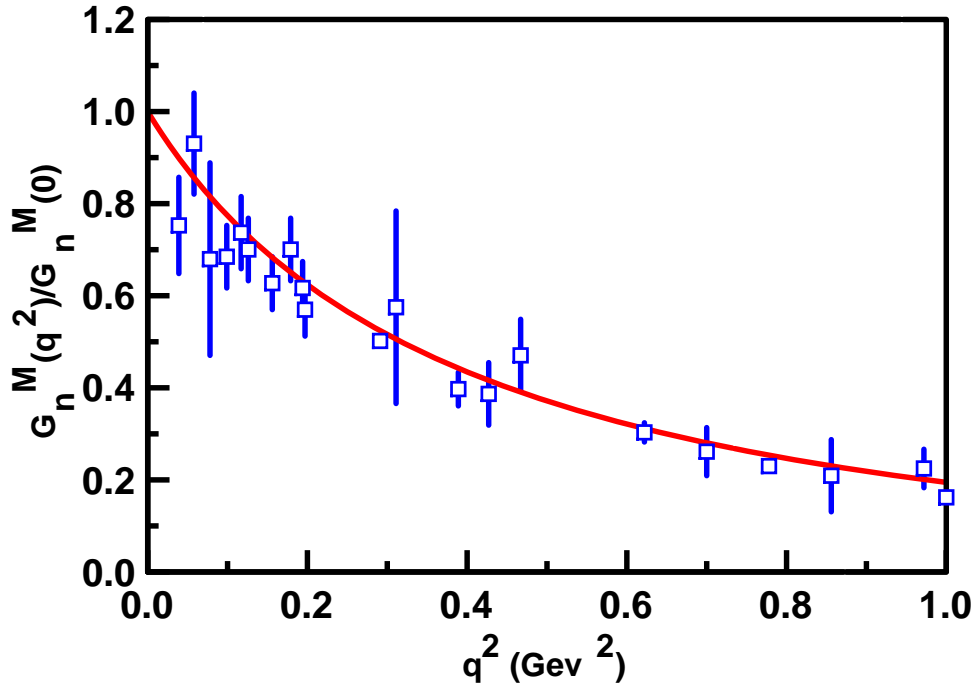}	\hfill}
\caption{Neutron magnetic form factor as a function of $q^2$ in $GeV^2$
calculated for $F_\pi = 93$, $e=2\pi$, $N_f=2$, $C_g=(300 MeV)^4$, and 
$m_\pi=139$.  The experimental data shown come from} Ref.~\cite{Bartel}
\label{NeutronM}
\end{figure}
The isoscalar part of the Skyrmion electric charge coincides with the
baryon density distribution, and for the isovector density from the
Skyrmion model one obtains, 
        \begin{eqnarray}
        \rho^{V}(x) = sin^2F(x)\left[x^2\rho^2(x) +
        \left(F^\prime(x)\right)^2x^2 + sin^2F(x)\right]\ .
        \label{c17}\end{eqnarray}

To take chiral symmetry breaking into account, we must add the pion mass
term,
        \begin{eqnarray}
        {\cal L}_\pi = {1\over 4} m_\pi^2 F_\pi^2 e^{-3\sigma}
        Tr\left[U + U^+ - {3\over 2} e^{-\sigma}\right]\ ,
        \label{c19}\end{eqnarray}
to our Skyrme model Lagrangian.  The theoretical predictions for the 
proton electric, neutron electric, proton magnetic and neutron magnetic
form factors, compared with data are shown in Figs. 
\ref{Proton}, \ref{Neutron}, \ref{ProtonM}, and \ref{NeutronM}, respectively.
Corresponding values of the proton and neutron mean square radius of the 
electric charge distribution are 0.78 $Fm^2$ and - 0.19 $Fm^2$.
 
\section{Conclusions}
\indent

We have presented our calculations on the nucleon electromagnetic form
factors in the framework of the generalized Skyrme model with dilaton
quarkonium field. The first calculation~\cite{NuovoCimento} in 
such a model yielded large deviations of the calculated form factors
from the dipole approximation formula.  In the present work, we use the
empirical value of the pion decay constant and the theoretical value for
the Skyrme term constant in the vector meson dominance approach to 
obtain a good description of the form factor data in the finite
range of momentum transfer in the measurements. The vector mesons
are included only as elements of the hadron substructure of the photon
and are not considered as components of the structure in the soliton
self-dynamics.  Implicit in the approach, though not explicitly
proposed, is the possibility of having the role of vector mesons given
by higher derivative terms in the effective Lagrangian for soliton
dynamics~\cite{Rajat}.  For example, keeping terms to four
orders in the expansion of the effective Lagrangian would lead to 
a $\rho$ meson-like term and the sixth order terms would give 
 $\omega$-like terms which are important in the calculations
of the form factors at the larger momentum transfers.


\begin{thebibliography}{18}
\bibitem{NuovoCimento}V.Nikolaev, O.Tkachev, V.Novozhilov,
        	      IL NUOVO CIMENTO, 1994, 107A(12), 2637.
\bibitem{Andrianov2}  Andrianov A.A., Andrianov V.A., Novozhilov Yu.V.,
                      Novozhilov V.Yu.
                      Phys.Lett. 1987, B186(3,4), 401.
\bibitem{Riska88}     Riska D.O. and Schwesinger B., 
                      Phys.Lett., 1989, B229, 339.
\bibitem{Gomm1}       Gomm M., Jain P., Jonson R. and Schehter J.
                      Phys.Rev. 1986, D33(3), 801.
\bibitem{Gomm2}       Gomm M., Jain P., Jonson R. and Schehter J.
                      Phys. Rev. 1986, D33(11), 3476.
\bibitem{Jain}        Jain P., Jonson R. and Schechter J.
                      Phys.Rev. 1987, D35(7), 2230.
\bibitem{Andrianov3}  Andrianov A.A., Novozhilov Yu.V. 
		      Phys.Lett. 1988, B202, 580.
\bibitem{Lacombe}     Lacombe M. Loiseau B., Vinh Mau, Cottingham W.N.
                      IPN Orsay preprint 87-28, 1987.
\bibitem{Yabu}        Yabu H., Schwesinger B., Holzwarth G.
                      Phys.Lett., 1989, B224, 25.
\bibitem{Mashaal}     Mashaal M., Phom  T.N., Truong T.N. 
                      Phys.Rev., 1986, D34, 3484.
\bibitem{Novikov0}    Novikov V.A., Shifman  M.A.,  Vainstein  A.,
                      Voloshin  M.B., Zakharov V.I.,
                      Nucl.Phys., 1984, B237, 525.
\bibitem{Nikolaev109} Nikolaev V.A. "Skyrme model and nucleon structure",
                      Dubna, Proceedings of the IX ISEPP, 1988, vol.1, 51.
\bibitem{Witten}      Adkins G.S., Nappi C.R. and Witten E.
		      Nucl.Phys. 1983, B228(3), 552.
\bibitem{Bartel}      Bartel W. et al., Nucl.Phys., 1973, B58, 429;
		      Kirk P.N. et al. Phys.Rev., 1973, D8, 63;
		      Hanson K. et al., Phys.Rev., 1973, D8, 753;
		      Rock S. et al., Phys.Rev.Lett., 1982, 49, 1139;
		      Arnold R.G. et al., Phys.Rev.Lett., 1986, 57, 174;
		      Bosted P.E. et al., Phys.Rev., 1990, C42, 38.
\bibitem{HOLZ}        Holzwarth G.,
		      Contribution to the Sixth International Symposium on 
		      Meson-Nucleon Physics and the Structure of the Nucleon, 
		      Blaubeuren/Tuebingen, Germany, 10-14 July, 1995. 
\bibitem{MEIER}	      Meier F., Walliser H.,
		      Physics Reports.,1997,289,p.448.
\bibitem{Rajat}	      Rajat K. Bhaduri, Lecture Notes and Supplements in
		      Physics:  Models of Nucleon. Addison-Wesley 
		      Publishing Company, Inc. 1988, p.283.
\end{thebibliography}
\end{document}